\documentclass{article}
\usepackage{spconf,amsmath,graphicx,hyperref}
\usepackage{cite}
\usepackage{amsmath,amssymb,amsfonts}
\usepackage{algorithmic}
\usepackage{graphicx}
\usepackage{textcomp}
\usepackage{xcolor}
\usepackage{setspace}
\usepackage{subcaption}
\usepackage{enumerate}
\usepackage{hyperref}
\usepackage{booktabs}
\usepackage{multirow}
\usepackage{enumitem}
\usepackage{mathrsfs}
\usepackage{enumerate}
\usepackage{bigstrut}
\usepackage{siunitx}
\usepackage{circledsteps}

\usepackage{caption}
\def\H{{\mathsf H}}

\def\CC{{\mathbb C}}
\def\RR{{\mathbb R}}

\usepackage{flushend}

\title{Why Not Put a Microphone Near the Loudspeaker?\\A New Paradigm for Acoustic Echo Cancellation }

\name{Fei Zhao$^{1,2}$ and Zhong-Qiu Wang$^2$
\thanks{This work was done while Fei Zhao was a visiting student at SUSTech. 
\textit{Corresponding author: Zhong-Qiu Wang}.}
}

\address{ 
  $^1$Inner Mongolia University, China\\
  $^2$Southern University of Science and Technology, China \\
\small \texttt{zhaofei@mail.imu.edu.cn, wang.zhongqiu41@gmail.com}
}

\begin{document}

% \ninept
\small
\setstretch{0.94}

\maketitle
\begin{abstract}
% AEC任务，非线性一直是个挑战，我们提出一种额外参考麦克风，为了解决参考麦克风中额外的近端信号，采用mask ref，接着用STWS，实验结果表明在match测试效果好，在强非线性这种unmatch的效果更突出。
Acoustic echo cancellation (AEC) remains challenging in real-world environments due to nonlinear distortions caused by low-cost loudspeakers and complex room acoustics. To mitigate these issues, we introduce a dual-microphone configuration, where an auxiliary reference microphone is placed near the loudspeaker to capture the nonlinearly distorted far-end signal. Although this reference signal is contaminated by near-end speech, we propose a preprocessing module based on Wiener filtering to estimate a compressed time-frequency mask to suppress near-end components.
This purified reference signal enables a more effective linear AEC stage, whose residual error signal is then fed to a deep neural network for joint residual echo and noise suppression. Evaluation results show that our method outperforms baseline approaches on matched test sets. To evaluate its robustness under strong nonlinearities, we further test it on a mismatched dataset and observe that it achieves substantial performance gains.
These results demonstrate its effectiveness in practical scenarios where the nonlinear distortions are typically unknown.

\end{abstract}

\section{Introduction}
% 第一段介绍AEC
% 第二段说AEC广泛存在的非线性问题，以及解决方案

Acoustic echo cancellation (AEC) is a critical signal processing task that aims to remove undesired echo signals caused by acoustic coupling between loudspeakers and microphones in voice communication systems \cite{benesty2001advances, hansler2005acoustic}. Traditional AEC methods primarily rely on adaptive filtering \cite{borrallo1992implementation, bershad1979analysis, kuech2014state}, such as NLMS and PBFDAF, which achieve robust performance across varying acoustic conditions through online adaptation. However, these linear models are inherently limited in handling nonlinear distortions \cite{guerin2004nonlinear, wada2011enhancement}, especially when low-cost audio hardware introduces harmonic artifacts or clipping. This limitation leads to persistent residual echo, degrading user experience in real-world deployments.

To address these limitations, deep learning-based AEC methods have emerged and can be broadly categorized into three paradigms:
\textbf{1) End-to-End Deep Neural Networks (DNN)}: These models learn a direct mapping from microphone signals to echo-canceled outputs using strong deep architectures \cite{zhao2024sdaec, zhang2022multi, DBLP:journals/taslp/ZhangLL023, Wang2022TFGridNetJournal}. While achieving state-of-the-art performance on in-distribution data, they exhibit limited generalizability to unseen acoustic environments and nonlinear distortions \cite{9413585, 9746272, Panchapagesan2022}. 
\textbf{2) Cascaded Traditional-Neural Systems}: This hybrid approach first applies conventional adaptive filtering for coarse echo suppression, followed by DNNs for residual echo removal \cite{sridhar2021icassp, cutler2024icassp, chen2023progressive, wang2021weighted}. Although it reduces training complexity and improves overall performance, the DNN's generalizability remains a bottleneck.
\textbf{3) Neural-Adaptive Hybrid Methods}, where DNNs are trained to predict the parameters for adaptive filters \cite{revach2021kalmannet, yang2023low, zhang2023kalmannet} (e.g., step sizes, regularization weights). Such methods inherit the generalization benefits of traditional AEC but often underperform purely data-driven or cascaded systems in terms of absolute echo suppression \cite{yang2023low}.

\begin{figure}[t]
        \centering
        \vspace{-0.1cm}
	\includegraphics[width=0.9\linewidth]{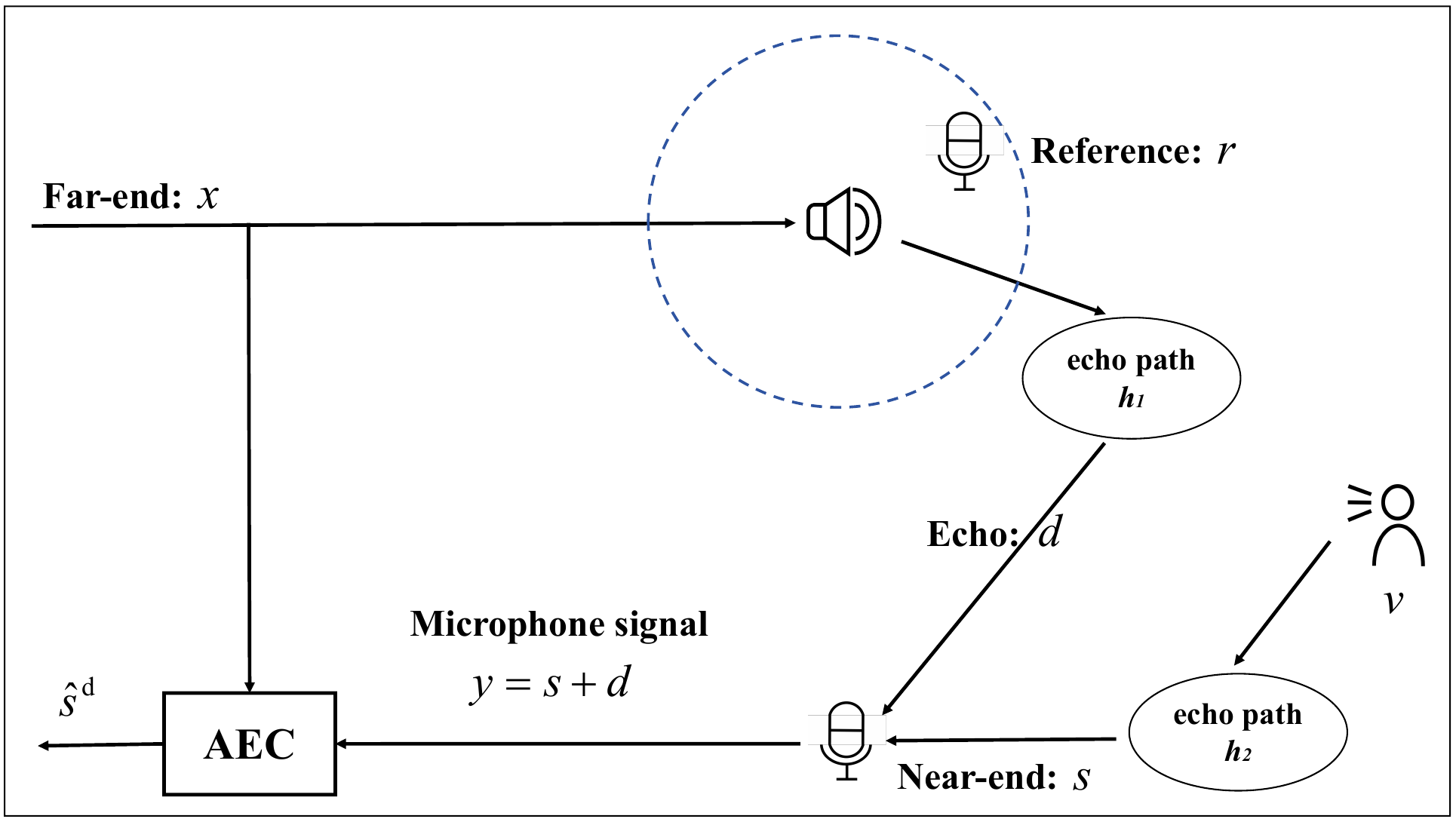}
	\caption{Illustration of AEC aided by using an auxiliary reference microphone.}
        \vspace{-0.5cm}%%减小图片上间隔
	\label{fig:another mic}
\end{figure}

% 第三段我们提出的额外麦克风解决方案
%   额外麦克风本身解决非线性问题
%   额外麦克风中有近端干扰，我们使用stws计算mask，得到优化的ref
%   所有的参考信号都做stws

% 可以大致说一下造成非线性的原因。扬声器与功放的非线性特性：包括硬剪辑（信号幅度超限截断）、Sigmoid类响应及功放饱和效应，导致谐波失真和互调失真。同时，当扬声器通常需要放大声音，这回造成扬声器过度震动，这种非正常的机械振动也会引起信号的非线性失真。

Nonlinear distortions in AEC systems typically originate from loudspeaker and amplifier characteristics, such as hard clipping, sigmoid-like amplitude responses, and amplifier saturation. These phenomena introduce harmonic and inter-modulation distortion. Additionally, excessive mechanical vibrations from high-intensity signals can induce further nonlinearities due to structural resonance or driver instability \cite{klippel2006tutorial}.

To address the prominent non-linearity challenge in AEC, we propose a microphone-augmented solution. As shown in Fig. \ref{fig:another mic}, an auxiliary reference microphone is placed near the loudspeaker in a conventional AEC system to capture nonlinearly distorted far-end reference signal. Prior work has explored multi-microphone AEC \cite{DBLP:journals/taslp/ZhangLL023}, but typically uses distant compact microphone arrays for spatial filtering. In contrast, the recorded reference signal contains the non-linearly distorted far-end signal, and could mitigate the difficulties in AEC caused by unknown non-linear distortions.
However, the reference signal also contains non-negligible near-end signals, which would interfere with nonlinearity-focused processing. To suppress the near-end signals, we employ a two-stage preprocessing pipeline: a short-time Wiener filter is first applied between the far-end and reference signals to estimate the far-end components, followed by computing an ideal ratio mask (IRM) to suppress the near-end components.
This yields a purified reference signal. Subsequently, the far-end signal, original reference signal, and purified reference signal are jointly processed with the main microphone signal using Wiener filtering for preliminary echo suppression. Finally, all signals are fed into the AEC model to predict the near-end direct signal.
Our evaluation results show that this method not only achieves significant performance gains over baseline systems but also exhibits superior robustness, particularly under strong, unseen nonlinear distortions such as clipping and amplifier saturation, confirming its effectiveness in practical scenarios.

\section{Problem Formulation}

Fig. \ref{fig:another mic} shows our proposed single-channel AEC system, which features an auxiliary reference microphone positioned in proximity to the loudspeaker. This strategic placement allows it to effectively capture the loudspeaker's nonlinear output. Assuming weak ambient noise, the mixture signal captured by the main microphone, $y$, is modeled as a mixture of the acoustic echo signal $d$ and near-end speech signal $s$.
The physical model in the time domain can be expressed as
\begin{align}
    y &= v * h_1 + x_{\text{nl}} * h_2 \\
    &= s + d \\ 
    &= s^\text{d} + s^\text{r} + d \in \RR^N,
    \label{1}
\end{align}
where $*$ denotes linear convolution and $N$ signal length in samples. The near-end signal $s$ arises from convolving the anechoic near-end source $v$ with its propagation room impulse response (RIR) $h_1$, and decomposes into two physically interpretable components: the direct-path component $s^\text{d}$ (unreflected sound from the near-end source to the main microphone) and late reverberant component $s^\text{r}$ (multiple late reflections of $v$ within the room). The acoustic echo $d$, meanwhile, is generated by convolving the loudspeaker's nonlinear output $x_{\text{nl}}$ (a distorted version of the far-end signal $x$) with the echo-path RIR $h_2$.

The mixture signal captured by the auxiliary reference microphone, $r$, is a composite of the far-end signal component, $r^{\text{f}}$, and the contaminating near-end signal, $r^{\text{n}}$, formulated as:
\begin{align}
    r &= v * h_3 + x_{\text{nl}} * h_4 \\
    &= r^{\text{n}} + r^{\text{f}}, 
\end{align}
where $h_3$ and $h_4$ respectively denote the RIRs from the loudspeaker and near-end talker to the reference microphone. Crucially, due to this microphone's proximity to the loudspeaker, the far-end component exhibits much larger energy than the near-end component.
This large energy disparity is the key physical property we exploit to isolate the nonlinear far-end signal for more effective AEC.

\section{Proposed method}

%\subsection{Overview}
Our proposed system, shown in Fig. \ref{fig: AEC Model}, operates in the short-time Fourier transform (STFT) domain.
It comprises two stages: a linear AEC stage, which suppresses the near-end component in the reference signal and performs signal processing-based linear AEC, and a neural AEC stage suppresses residual echoes.

%\subsection{Linear AEC and Reference Signal Preprocessing}
\subsection{Linear AEC Stage}

\subsubsection{Weighted short-time Wiener solution}

% 对于线性AEC方法我们采用加权的短时维纳解（WSTWS），该方法是对短时维纳解方法的改进，通过加权来平衡不同时频单元(T-F units)对线性预测的贡献。具体公式如下：
For the linear AEC component, we adopt the weighted short-time Wiener solution (WSTWS) \cite{wang2021convolutive}, an enhancement over the conventional short-time Wiener solution (STWS) \cite{zhao2025attention} by adaptively re-weighting the contribution of individual time-frequency (T-F) units \cite{wang2021convolutive}.
The linear AEC result is obtained as 
\begin{equation}
    \mathcal{F}_{\text{WSTWS}}(Y, X) = Y(t, f) - \hat{\mathbf{h}}_1(t,f)^{\mathrm{H}}\mathbf{X}(t, f),\label{FCP_result}
\end{equation}
which represents the result of canceling the estimated echo signal from the near-end mixture signal $Y$ based on the far-end signal $X$.
In Eq. (\ref{FCP_result}), $\hat{\mathbf{h}}_1(t,f) \in \CC^K$ denotes a $K$-tap linear filter modeling the echo path, and $\mathbf{X}(t, f) = [X(t,f), X(t-1, f), ..., X(t-K+1, f)]^\mathsf{T} \in \CC^K$.
In frame-online, real-time AEC, at the current frame $t$, WSTWS estimates the filter by solving the following problem:
\begin{equation}
    \hat{\mathbf{h}}_1(t,f) = \underset{\mathbf{h_1}(f)}{\operatorname{argmin}}\sum_{t'=t-W}^{t}\frac{\left|{Y}(t', f)-\mathbf{h}_1(f)^{\H}\mathbf{X}(t', f)\right|^{2}}{{\lambda}(t, f)}, \label{FCP_optimization}
\end{equation}
which is computed based on the current and the past $W$ frames.
%To satisfy real-time processing constraints, the solution at time $t$ is computed causally using a $W$-frame historical window.
The weighting term $\lambda$ is computed as
\begin{equation}
\lambda(t, f)=\varepsilon \times \max_{t' \in [t-W,t]} \left(|Y(t', f)|^{2}\right)+|Y(t, f)|^{2}, \label{weighting_term}
\end{equation}
where $\varepsilon$ is a floor value to avoid placing too much weight on low-energy T-F units.

\subsubsection{Reference signal preprocessing}
\label{rsp}

Although the reference signal contains strong energy of the far-end signal, it still contains a significant amount of near-end signals.
To minimize the interference of the near-end components in the reference signal, we first leverage WSTWS to estimate the near-end components and then employ a method similar to real-valued T-F masking \cite{narayanan2013ideal} to suppress the near-end components.
In detail,
\begin{equation}
\hat{M} = \frac{|R-\mathcal{F}_{\text{WSTWS}}(R, X)|}{|R-\mathcal{F}_{\text{WSTWS}}(R, X)| + |\mathcal{F}_{\text{WSTWS}}(R, X)|},
\label{ref_mask}
\end{equation}
\begin{equation}
R_m = \hat{M}^m \odot R,
\label{masked_ref}
\end{equation}
where $\mathcal{F}_{\text{WSTWS}}(R, X)$ is the WSTWS result calculated based on the reference signal and the far-end signal, $m\in[0,1]$ is a tunable mask compression factor, and $\odot$ denotes point-wise multiplication.
Because the far-end component in the reference signal has dominant energy, Eq. (\ref{ref_mask}) preserves this dominant far-end component.
%Furthermore, the compressed mask in Eq. (\ref{masked_ref}) is applied to the reference signal to suppress the near-end component, where $m$ is the compression factor.
%At this time, two additional linear filtering results can be obtained: $\mathcal{F}_{\text{WSTWS}}(Y, R)$ and $\mathcal{F}_{\text{WSTWS}}(Y, R_m)$.
With $R_m$, we can compute an additional linear AEC result, denoted as $\mathcal{F}_{\text{WSTWS}}(Y, R_m)$, besides using the reference signal $R$ to directly compute a linear AEC result (i.e., $\mathcal{F}_{\text{WSTWS}}(Y, R)$).

\subsection{Neural AEC Stage}
\label{Am}

% 分两块 一块是网络主体，另一个是信息预处理
\begin{figure}[t]
        \centering
        \vspace{-0.1cm}
	\includegraphics[width=0.9\linewidth]{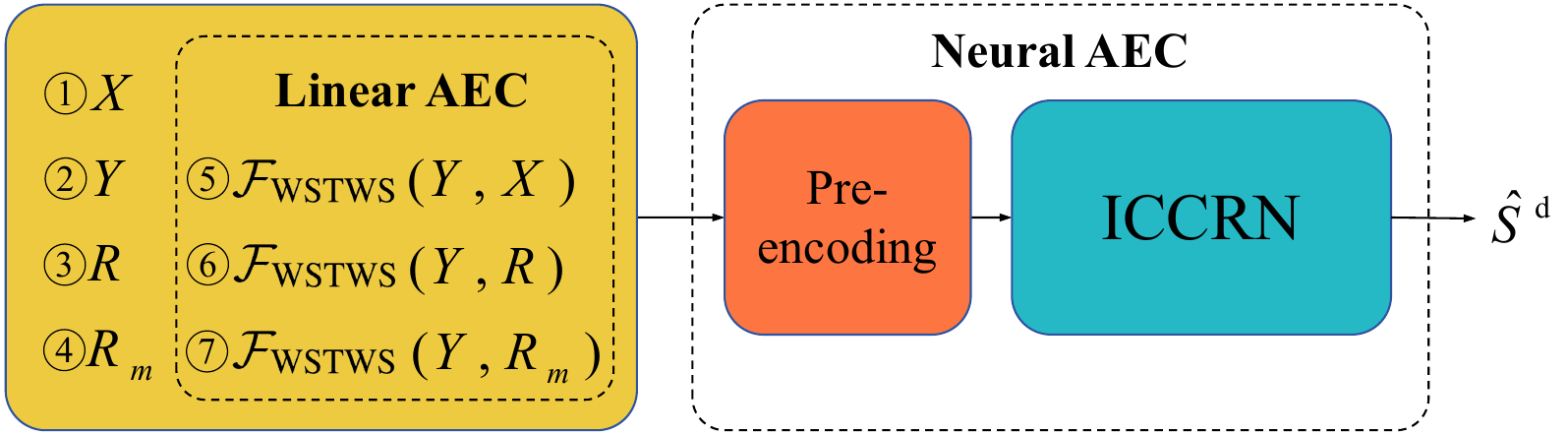}
	\caption{Architecture of proposed AEC system.
    %: Comprehensive Diagram.
    }
        % \vspace{-0.6cm}%%减小图片上间隔
	\label{fig: AEC Model}
            \vspace{-0.5cm}
\end{figure}

%the T-F representations of 
Fig. \ref{fig: AEC Model} illustrates the overall architecture of the proposed AEC model. The network takes $7$ input signals: \Circled{\footnotesize 1} far-end signal $X$, \Circled{\footnotesize 2} near-end mixture signal $Y$, and \Circled{\footnotesize 3} reference signal $R$ and \Circled{\footnotesize 4} its masked version $R_m$.
Additionally, we include the outputs of WSTWS applied to three signal pairs: \Circled{\footnotesize 5}  $\mathcal{F}_{\text{WSTWS}}(Y, X)$, \Circled{\footnotesize 6} $\mathcal{F}_{\text{WSTWS}}(Y, R)$, and \Circled{\footnotesize 7} $\mathcal{F}_{\text{WSTWS}}(Y, R_m)$, which respectively represent the linear AEC results between the far-end signal, the reference signal, the masked reference signal, and the near-end mixture signal. This multi-input design enables the model to leverage complementary echo estimates to improve AEC performance. 

To harmonize the heterogeneous input representations and align their temporal dynamics, we apply a dedicated pre-encoding block to the real and imaginary components of each of the seven input signals. Each pre-encoding block consists of a causal 2D convolutional layer with a $3\times 3$ kernel, $2$ input channels and $2$ output channels, followed by layer normalization \cite{ba2016layer} and PReLU \cite{he2015delving}.
The processed features are then concatenated along the channel dimension, resulting in a $14$-channel input tensor to the AEC model.
%—corresponding to the real and imaginary components of the seven pre-encoded signals, to meet the input requirements of the downstream network.
The AEC model employs the ICCRN architecture \cite{DBLP:conf/icassp/LiuZ23d}.
It is trained to output an estimate of the near-end speech signal, denoted as $\hat{S^d}$.

\subsection{Loss Functions}

The loss function combines the two loss functions:
\begin{equation}
	\mathcal{L}_{\text{total}} = \mathcal{L}_{\text{RI+Mag}} + \alpha \times \mathcal{L}_{\text{S-SISNR}},
\end{equation}
where $\alpha$ is tuned to $0.01$.
The first loss term, $\mathcal{L}_{\text{RI+Mag}}$, penalizes the complex spectrum as follows:
\begin{equation}
	\mathcal{L}_{\text{RI+Mag}} = \mathcal{L}_{\mathrm{RI}} + \mathcal{L}_{\text{Mag}},
\end{equation}
\begin{equation}
	\mathcal{L}_{\mathrm{RI}} = \sum_{t,f} \Big| |S^{\text{d}}(t, f)|^p e^{j \angle{S^{\text{d}}(t, f)}}-|\hat{S}^{\text{d}}(t, f)|^p e^{j \angle{\hat{S}^{\text{d}}(t, f)}} \Big|^2,
\end{equation}
\begin{equation}
	\mathcal{L}_{\mathrm{Mag}}=\sum_{t,f} \Big||S^{\text{d}}(t, f)|^p-|\hat{S}^{\text{d}}(t, f)|^p\Big|^2 ,
\end{equation}
where $p$ (tuned to $0.5$) is a spectral compression factor following \cite{li2021importance}.
% Operator $\theta$ calculates the phase of a complex number.
The second loss term, $\mathcal{L}_{\text{S-SISNR}}$, denotes the stretched scale-invariant signal-to-noise ratio (S-SISNR) loss \cite{DBLP:conf/interspeech/SunYZH21}, which is a variant of SISNR \cite{DBLP:journals/taslp/LuoM19}. 
It is a time domain loss function obtained by doubling the period of SISNR:
%. The simplified formula for S-SISNR is expressed as follows:
\begin{equation}
	\mathcal{L}_{\text {S-SISNR }}=10 \log_{10} cot^2(\frac{\beta}{2}) = 10\log_{10}\frac{1+cos(\beta)}{1-cos(\beta)},
\end{equation}
where $cos(\beta)$ denotes the cosine similarity between the target speech signal $s^d$ and predicted speech signal $\hat{s}^d$.
%, since it is complicated to calculate the half angle, after the derivation of the trigonometric function, it can be represented by $cos(\beta)$.

\section{Experimental Setup}
\subsection{Datasets}

The near-end and far-end signals employed in our experiments are sourced from the synthetic datasets of the ICASSP $2023$ AEC challenge \cite{cutler2024icassp} to simulate the near-end and far-end scenarios.
The RIRs are generated by using the image method \cite{allen1979image}. 
We simulate different rooms of dimensions $l \times w \times h$, with $l$ randomly-sampled from $4$ to $8$ meters, $w$ from $3$ to $7$ meters, and $h$ from $3$ to $5$ meters. The positions of the sound sources (including the loudspeaker and the near-end talker) are randomly set within the room.
The auxiliary reference microphones are randomly set within a hollow sphere with a radius of $0.05$ to $0.2$ meters, centered at the loudspeaker position. The main microphones are randomly placed in the cavity with a length range of $[l/10, l-1/10]$, a width range of $[w/10, w-w/10]$, and a height range of $[1, min(h-1, 3)]$.
The reverberation time (T$60$) is randomly sampled from $0.1$ to $0.8$ seconds
%for simulating rooms
to cover a wide range of acoustic conditions.
In total, $22,000$ RIRs are generated based on the aforementioned configuration, and $20,000$ of them are allocated for training, $1,000$ for validation, and $1,000$ for testing.
The generated echo signal is mixed with near-end speech $s$ at a signal-to-echo ratio (SER) randomly sampled from the range $[- 10, 10]$ dB with a step of $1$ dB.
We generate a dataset consisting of $100,000$ training, $5,000$ validation, and $1,000$ test mixtures.
All samples are $6$-second long.
The sampling rate is $16$ kHz.

\subsection{Nonlinearity Scenarios}

This study considers five types of nonlinear settings.
The first three \cite{DBLP:conf/icassp/ZhangLZ22} are used for training.
% For testing, we construct a matched test condition that also uses the first three for testing, and a mismatched test condition that uses the last two for testing to verify the effectiveness of introducing auxiliary reference microphones.
For testing, we construct two test conditions: a matched condition, which employs the first three nonlinear settings, and a mismatched condition, which employs the last two, to verify the effectiveness of introducing auxiliary reference microphones.
\begin{align*}
x_{nl}(n) &= \frac{a x(n)}{\sqrt{a^2 + x^2(n)}}, \text{with}\,a = \frac{5}{b},\tag{16} \\
x_{nl}(n) &= 1 - e^{-a x(n)}, \text{with}\,a = \frac{b}{10} ,\tag{17} \\
x_{nl}(n) &= 2ax(n) + ax^2(n) + x^3(n), \text{with}\,a = \log\big(\frac{b}{10}\big) + 0.1,  \tag{18} 
\end{align*}
where $b \in [2,5]$. The other two nonlinearities are respectively the combinations of hard-clipping \cite{stenger2000adaptation} and soft-clipping \cite{nollett1997nonlinear} with sigmoid nonlinearity:
\begin{align*}
x_{\text{hard}}(n) &= 
\begin{cases} 
-x_{\text{max}}, & x(n) < -x_{\text{max}} \\
x(n), & |x(n)| \leq x_{\text{max}} \\
x_{\text{max}}, & x(n) > x_{\text{max}}
\end{cases} \tag{19} \\
x_{\text{soft}}(n) &= \frac{x(n)x_{\text{max}}}
{\sqrt{|x_{\text{max}}|^{\rho} + |x(n)|^{\rho}}}, \tag{20}
\end{align*}
where the clipping threshold $x_{\text{max}}$ is set to $0.7$ in this work.
For soft-clipping saturation distortion, $\rho$ is set to $2$, denoting the
non-adaptive shape parameter.
\begin{equation}
    x_{\text{sigmoid}}(n) = \lambda \left( \frac{1}{1 + e^{-\nu \cdot z(n)}} - \frac{1}{2} \right), \tag{21}
\end{equation}
where $z(n) = 1.5 \times x(n) - 0.3 \times x^2(n)$, 
the parameter $\lambda$ denotes the sigmoid gain and is set to 2, and $\nu$ represents the sigmoid slope and is defined as
\begin{align*}
\nu = 
\begin{cases} 
4, & \text{if } z(k) > 0 \\
0.5, & \text{if } z(k) \leq 0  \tag{22} 
\end{cases}
\end{align*}
For the unmatched test set, the same sound source signals as in the matched test set are used, but with an invisible non-linear setting.

\subsection{Training Details}

For STFT, the window size is $20$ ms, hop size $10$ ms, and the Hamming window is used as the analysis window. 
% Given $16$ kHz sampling rate, a $320$-point discrete Fourier transform (DFT) is computed.
The model is trained using the Adam optimizer, starting with an initial learning rate of $0.001$, which is halved if the validation loss does not improve for $2$ epochs.
Early stopping is applied if no improvement is observed for $8$ epochs.
The number of filter taps in WSTWS is tuned to $K=20$ in default, while in Eq. (\ref{ref_mask}),
%since the far-end component in the reference signal is mainly the direct signal,
considering that the loudspeaker and the reference microphone are spatially close to each other,
the number of filter taps for WSTWS is tuned to $1$.
in Eq. (\ref{weighting_term}), the flooring term $\varepsilon$ is tuned to $0.001$.
In Eq. (\ref{FCP_optimization}), the window size $W$ is tuned to $200$ frames.
In Eq. (\ref{masked_ref}), the mask compression factor $m$ is tuned $1/6$.

\subsection{Evaluation Setup}
\label{sec:Evaluation setup}

We consider three evaluation scenarios, including double talk (DT), near-end single talk (ST\_NE), and far-end single talk (ST\_FE).
Three popular evaluation metrics are used, including echo return loss enhancement (ERLE) \cite{enzner2014acoustic} for measuring echo suppression during far-end single talk, perceptual evaluation of speech quality (PESQ) \cite{DBLP:conf/icassp/RixBHH01} for assessing near-end speech quality, and signal-to-distortion ratio (SDR) \cite{vincent2006performance} for quantifying near-end speech fidelity during double talk.
The evaluated models consist of linear AEC methods, ICCRN, and ICCRN$_*$. Specifically, the linear AEC methods include $\mathcal{F}_{\text{WSTWS}}(Y, X)$, $\mathcal{F}_{\text{STWS}}(Y, R_m)$, and $\mathcal{F}_{\text{WSTWS}}(Y, R_m)$, 
while ICCRN$_*$ denotes variants that utilize different input features.
In detail, 
\begin{itemize}[leftmargin=*,noitemsep,topsep=0pt]
\item \textbf{ICCRN} uses only the far-end signal and the near-end mixture signal as inputs to predict the target speech.

\item \textbf{ICCRN$_\text{r}$}, which builds on ICCRN, additionally incorporates the reference signal into its network input.

\item \textbf{ICCRN$_\text{rl}$}, building on \textbf{ICCRN$_\text{r}$}, further takes the linear AEC result computed based on the reference signal and the near-end mixture signal as input.
It should be noted that the linear AEC result used here is computed based on STWS. That is, $\mathcal{F}_{\text{STWS}}(Y, R)$.

\item \textbf{ICCRN$_\text{rl+fl}$}, building on \textbf{ICCRN$_\text{rl}$}, further incorporates the STWS result of the near-end mixture signal and the far-end signal. That is, $\mathcal{F}_{\text{STWS}}(Y, X)$.

\item \textbf{ICCRN$_\text{rl+fl+mrl}$}, building on \textbf{ICCRN$_\text{rl+fl}$}, additionally includes the masked reference signal $R_m$ in Section \ref{rsp} and $\mathcal{F}_{\text{STWS}}(Y, R_m)$ to the input signal.

\item \textbf{ICCRN$_\text{rl+fl+mrl+w}$}, building on \textbf{ICCRN$_\text{rl+fl+mrl}$}, replaces all the STWS operations with WSTSW. The detailed input signals are listed in Section \ref{Am}.
\end{itemize}
%Notice that ICCRN and ICCRN$_\text{rl}$ are considered as the baselines of the other models, as they are currently the popular way to realize neural AEC.
Notice that ICCRN is considered the baseline of the other models, as this is currently the popular way to realize neural AEC.

% baseline and proposed

\vspace{-0.05cm}
\section{Evaluation Results}
\label{sec:results}

\begin{table}[]
\footnotesize
  \centering
  \caption{AEC results on the matched test set.
  %(best scores are in bold).
  }
          \vspace{-0.25cm}
  \sisetup{table-format=2.2,round-mode=places,round-precision=2,table-number-alignment = center,detect-weight=true,detect-inline-weight=math}
  \setlength{\tabcolsep}{3.5pt}
  \scalebox{1.0}{
\fontsize{8}{10}\selectfont%设置字体大小
    \begin{tabular}{
    l
    S[table-format=1.2,round-precision=2] % 
    S[table-format=2.1,round-precision=1] % 
    S[table-format=1.2,round-precision=2] % 
    S[table-format=2.1,round-precision=1] % 
    S[table-format=2.1,round-precision=1] % 
    }
    \toprule
    Test scenarios & \multicolumn{2}{c}{DT}         & \multicolumn{2}{c}{ST\_NE}        & \multicolumn{1}{l}{ST\_FE} \\
    %\textbackslash{}Metric
    \cmidrule(lr){2-3}\cmidrule(lr){4-5}\cmidrule(lr){6-6}
    Model & \multicolumn{1}{l}{PESQ} & \multicolumn{1}{l}{SDR (dB)} & \multicolumn{1}{l}{PESQ} & \multicolumn{1}{l}{SDR (dB)} & \multicolumn{1}{l}{ERLE (dB)} \\
    \midrule
    Mixture     & 1.56 & -11.8  & 2.25    & -36.1   & {-} \\
    \midrule 
    $\mathcal{F}_{\text{WSTWS}}(Y, X)$ &1.85 &-8.23 &{-} &{-} &12.35  \\
    $\mathcal{F}_{\text{STWS}}(Y, R_m)$ &1.76 &-7.04 &{-} &{-} &17.54  \\
    $\mathcal{F}_{\text{WSTWS}}(Y, R_m)$ &1.88 &-8.04  &{-} &{-} &13.42  \\
    \midrule
    ICCRN                            &2.34	&4.09	&2.92	&5.09	&73.50 \\
    ICCRN$_\text{r}$                 &2.37	&4.05	&2.93	&5.27	&74.81 \\
    ICCRN$_\text{rl}$                &2.49	&4.71	&2.96	&5.06	&73.38 \\
    ICCRN$_\text{rl+fl}$             &2.52	&4.68	&2.93	&5.25	&76.10 \\
   ICCRN$_\text{rl+fl+mrl}$          &2.55	&5.01	&2.95	&5.76	&77.15 \\
   \textbf{ICCRN$_\text{rl+fl+mrl+w}$}        &\bfseries 2.59	&\bfseries 5.14	&\bfseries 2.97	& \bfseries 5.86	& \bfseries 77.82 \\
    \bottomrule
    \end{tabular}%
    }
    \vspace{-0.35cm}
  \label{tab:test_match}%

\end{table}%

\begin{table}[]
  \centering
  \caption{AEC results on the mismatched test set.
  %(best scores are in bold).
  }
            \vspace{-0.25cm}

    \sisetup{table-format=2.2,round-mode=places,round-precision=2,table-number-alignment = center,detect-weight=true,detect-inline-weight=math}
      \setlength{\tabcolsep}{3.5pt}
  \scalebox{1.0}{
\fontsize{8}{10}\selectfont%设置字体大小
    \begin{tabular}{
    l
    S[table-format=1.2,round-precision=2] % 
    S[table-format=2.1,round-precision=1] % 
    S[table-format=1.2,round-precision=2] % 
    S[table-format=2.1,round-precision=1] % 
    S[table-format=2.1,round-precision=1] % 
    }
    \toprule
    Test scenarios & \multicolumn{2}{c}{DT}         & \multicolumn{2}{c}{ST\_NE}        & \multicolumn{1}{l}{ST\_FE} \\
    %\textbackslash{}Metric
    \cmidrule(lr){2-3}\cmidrule(lr){4-5}\cmidrule(lr){6-6}
    Model & \multicolumn{1}{l}{PESQ} & \multicolumn{1}{l}{SDR (dB)} & \multicolumn{1}{l}{PESQ} & \multicolumn{1}{l}{SDR (dB)} & \multicolumn{1}{l}{ERLE (dB)} \\
    \midrule
    Mixture     & 1.52    & -12.0  & 2.28    & -36.6   & {-} \\
    \midrule
    $\mathcal{F}_{\text{WSTWS}}(Y, X)$ &1.65 &-8.99  &{-} &{-} &7.71  \\
    $\mathcal{F}_{\text{STWS}}(Y, R_m)$ &1.73 &-7.11  &{-} &{-} &16.83 \\
    $\mathcal{F}_{\text{WSTWS}}(Y, R_m)$ &1.85 &-8.07  &{-} &{-} &13.36 \\
    \midrule
    ICCRN                            &2.05	   &3.1	    &2.92	  &5.1	   &36.2 \\
    ICCRN$_\text{r}$                 &2.23	   &3.7	    &2.93	  &5.3	   &\bfseries 74.0 \\
    ICCRN$_\text{rl}$                &2.10	   &3.4 	&2.96	  &5.1     &37.4 \\
    ICCRN$_\text{rl+fl}$             &2.29     &4.2 	& \bfseries 2.98	  &5.6	   &65.8 \\
   ICCRN$_\text{rl+fl+mrl}$          &2.38 	&4.3	 &2.93	   &5.6  	&70.0 \\
   \textbf{ICCRN$_\text{rl+fl+mrl+w}$}        &\bfseries 2.47 	&\bfseries 4.8	 &2.97     &\bfseries 5.9 	&72.7 \\
    \bottomrule
    \end{tabular}%
    }
    \vspace{-0.35cm}
  \label{tab:test_unmatch}%
\end{table}%

The experimental results on both matched and mismatched test sets demonstrate the effectiveness of the proposed systems.

In Table~\ref{tab:test_match}, we observe \textbf{ICCRN$_\text{rl+fl+mrl+w}$} achieves the best performance across all evaluation metrics on the matched test set.
For the DT scenario, it demonstrates a $10.7\%$ improvement in PESQ score ($2.59$ vs. $2.34$) and a $24.3\%$ increase in SDR ($5.1$ vs. $4.1$ dB) compared to the baseline ICCRN.
In the ST\_NE condition, it attains the highest PESQ score of $2.97$ and achieves $5.86$ dB SDR, representing a $15.7\%$ improvement over ICCRN.
For the ST\_FE scenario, it delivers superior echo suppression performance with an ERLE score of $77.8$ dB.
% Second, the progressive performance gains from ICCRN and ICCRN$_\text{r}$ to ICCRN$_\text{rl+fl+mrl+w}$ validate the effectiveness of (a) linear AEC integration ($\mathcal{F}_{\text{STWS}}$); (b) weighted short-time Wiener solution (denoted by $\mathcal{F}_{\text{WSTWS}}$); and (3) masked reference signal ($R_m$) processing. 

In Table \ref{tab:test_unmatch}, we observe that \textbf{ICCRN$_\text{rl+fl+mrl+w}$} maintains strong performance in domain-mismatched conditions, demonstrating significant improvements over the baseline ICCRN.
In the DT scenario, it achieves a $20.4\%$ higher PESQ score ($2.47$ vs. $2.05$) and a $54.8\%$ enhancement in SDR ($4.8$ vs. $3.1$ dB), highlighting superior speech quality and fidelity. For ST\_FE scenario, it sustains a high ERLE score of $72.7$ dB, outperforming most variants (except ICCRN$_\text{r}$ by $3.9\%$, which underscores its robust AEC capability even in mismatched conditions.

Introducing the reference signal (ICCRN$_\text{r}$ vs. ICCRN) improves ERLE by 37.8 dB (74.0 vs. 36.2 dB) on the mismatched test set, validating the utility of the nonlinear reference signal provided by the auxiliary microphone.
Comparisons between ICCRN and ICCRN$_\text{rl+fl}$ demonstrate the effectiveness of the linear AEC method STWS, which yields significant performance improvements across both matched and mismatched test conditions. Further comparisons between ICCRN$_\text{rl+fl}$ and ICCRN$_\text{rl+fl+mrl}$ reveal that linear AEC applied to the purified reference signal also results in notable performance gains. This improvement is particularly pronounced in nonlinear scenario evaluations, as evidenced by the quantitative score differences between $\mathcal{F}_{\text{WSTWS}}(Y, X)$ and $\mathcal{F}_{\text{WSTWS}}(Y, R_m)$. Finally, comparisons between ICCRN$_\text{rl+fl+mrl}$ and ICCRN$_\text{rl+fl+mrl+w}$, alongside those between $\mathcal{F}_{\text{STWS}}(Y, R_m)$ and $\mathcal{F}_{\text{WSTWS}}(Y, R_m)$, confirm that the WSTWS further enhances performance.
Collectively, these results validate the model’s strong generalizability and effectiveness in adapting to cross-domain acoustic environments.

\section{conclusion}

This paper addresses the challenge of nonlinear distortions in AEC, commonly caused by low-cost loudspeakers and amplifiers.
We propose a novel AEC framework using an auxiliary reference microphone placed near the loudspeaker to capture nonlinear far-end signals. To suppress near-end speech contamination in the reference microphone signal, we propose a short-time Wiener filtering technique followed by real-valued ratio masking to generate a masked, cleaner reference signal.
% The processed signals include the far-end, primary microphone, and masked reference are jointly utilized for echo cancellation.
% The far-end signal, the near-end microphone signal, and the masked reference signal are used as the original signals.
A weighted STWS module further enhances preliminary suppression before neural processing.
Evaluation results show that our method significantly outperforms baselines on both matched and mismatched test sets.
It achieves the best performance across all evaluation metrics in matched conditions, and even greater gains ($20.4\%$ PESQ and $54.8\%$ SDR) under strong, unseen nonlinearities in double talk scenario.
The results demonstrate superior robustness and generalization, validating the effectiveness of introducing an auxiliary reference microphone to help AEC and the STWS-based masking strategy.

{\footnotesize
\bibliographystyle{IEEEtran}
\bibliography{refs}
}

\end{document}